%% file: main.tex
\def\year{2021}\relax
\documentclass[letterpaper]{article} 
\usepackage{aaai21}  
\usepackage{times}  
\usepackage{helvet} 
\usepackage{courier}  
\usepackage[hyphens]{url}  
\usepackage{graphicx} 
\usepackage{natbib}  
\usepackage{caption} 
\usepackage{diagbox}
\usepackage{soul}
\usepackage{url}
\usepackage{subcaption}
\usepackage{graphicx}
\usepackage[utf8]{inputenc}
\usepackage{amsmath}
\usepackage{booktabs}
\usepackage{algorithm}
\usepackage{algorithmicx}
\usepackage{amsfonts,amssymb}
\usepackage{enumitem}
\usepackage{amsmath, amssymb}
\usepackage{makecell}
\usepackage{multirow}
\usepackage{algpseudocode}
\usepackage{bm}

\usepackage{color}
\usepackage{tabularx}
\usepackage[font=small,skip=0pt]{caption}
\usepackage{graphicx,subcaption}
\usepackage{varwidth}

\usepackage{xcolor}

\nocopyright

\title{RRCN: A Reinforced Random Convolutional Network based Reciprocal Recommendation Approach for Online Dating}
\author{

    Linhao Luo\textsuperscript{\rm 1}, Liqi Yang\textsuperscript{\rm 1}, Ju Xin\textsuperscript{\rm 1}, Yixiang Fang\textsuperscript{\rm 2}\\
     Xiaofeng Zhang\textsuperscript{\rm 1}, Xiaofei Yang\textsuperscript{\rm 3}, Kai Chen\textsuperscript{\rm 1}, Zhiyuan Zhang\textsuperscript{\rm 1}, Kai Liu\textsuperscript{\rm 4}\\
}
\affiliations{

    \textsuperscript{\rm 1} Harbin Institute of Technology, ShenZhen\\
    \textsuperscript{\rm 2} University of New South Wales\\
    \textsuperscript{\rm 3} Peng Cheng Laboratory\\
    \textsuperscript{\rm 4} Pingan\\
    luolinhao@stu.hit.edu.cn

}

\begin{document}

\maketitle

\input{abstract}

\input{introduction}

\input{preliminaries}

\input{approach}

\input{experiment}

\input{relatedwork}
\input{conclusion}

\bibliography{main}

\end{document}

%% file: abstract.tex
\begin{abstract}
Recently, the reciprocal recommendation, especially for online dating applications, has attracted more and more research attention. Different from conventional recommendation problems, the reciprocal recommendation aims to simultaneously best match users' mutual preferences.
Intuitively, the mutual preferences might be affected by a few key attributes that users like or dislike. Meanwhile, the interactions between users' attributes and their key attributes are also important for key attributes selection. Motivated by these observations, in this paper we propose a novel reinforced random convolutional network (RRCN) approach for the reciprocal recommendation task. In particular, we technically propose a novel random CNN component which can randomly convolute non-adjacent features to capture their interaction information and learn feature embeddings of key attributes to make the final recommendation. Moreover, we design a reinforcement learning based strategy to integrate with the random CNN component to select salient attributes to form the candidate set of key attributes. We evaluate the proposed RRCN against a number of both baselines and the state-of-the-art approaches on two real-world datasets, and the promising results have demonstrated the superiority of RRCN against the compared approaches in terms of a number of evaluation criteria.
\end{abstract}

%% file: introduction.tex
\section{Introduction}\label{sec:intro}
Nowadays, the most popular online dating Web applications could even have several hundreds of millions of registered users. Consequently, an effective reciprocal recommendation system \cite{neve2019latent,ting2016transfer,palomares2020reciprocal} is urgently needed to enhance user experience. Generally, the reciprocal recommendation problem aims to recommend a list of users to another user that best matches their mutual interests \cite{pizzato2013recommending,zheng2018fairness}. For example in an online dating platform (e.g., Zhenai \footnote{https://www.zhenai.com/} or Match \footnote{https://www.match.com/}), the purpose of reciprocal recommendation is to recommend male users and female users who are mutually interested in each other.

Generally, the online dating users and their historical messages can often be modeled as an attributed bipartite graph \cite{zhao2013user,zhang2017learning,sheikh2019gat2vec}, where nodes represent users, directed edges represent messages passing among users, and nodes are associated with some attributes. In the bipartite graph, there are two types of edges, i.e., reciprocal links and non-reciprocal links. A reciprocal link indicates that a user sent a message to and was replied by another user, whereas a non-reciprocal link means that a user sent a message to but was not replied by another user. Accordingly, the reciprocal recommendation problem could be cast into the reciprocal link prediction problem \cite{xia2014predicting}.

\textbf{Prior works.} In the literature, there are various recommendation approaches \cite{guo2017deepfm,lian2018xdeepfm,li2019multi,xi2019bpam,chen2019matching}. For example, DeepFM \cite{guo2017deepfm} and xDeepFM \cite{lian2018xdeepfm} are proposed with a focus on extracting the low- and high-order features as well as their interactions. However, these conventional recommendation approaches \cite{tang2013exploiting,davidson2010youtube,hicken2005music,wei2017collaborative} cannot be directly adapted to the reciprocal recommendation problem, since they only care the interest of one side. 
Recently, a few approaches \cite{nayak2010social,pizzato2010recon,chen2011recommendation,kleinerman2018optimally} have been proposed to address this issue. However, most of them convert this task to a two-stage conventional recommendation problem. For instance, RECON \cite{pizzato2010recon} measures mutual interests between a pair of users for reciprocal recommendation task. 
Unfortunately, these approaches mainly consider the effect of attributes of preferred users, but overlook the effect of attributes of disliked users. Last but not least, they treat all the attributes equally, which ignores the fact that different attributes may have different impacts on the reciprocal recommendation \cite{wang2013feature, boutemedjet2008unsupervised, zheng2012optimal}.

Intuitively \cite{hitsch2005makes,pizzato2010learning}, a user might send a message to another user if and only if the other user has certain content of profile that is preferred by the user, denoted as user's \textit{preferred attribute}. On the contrary, if a user does not reply to a message, it indicates that either there are no \textit{preferred attributes} or there is at least one attribute of the other user that the user does not like, which is called \textit{repulsive attribute} in this paper. For example, user A with a good \textit{salary} may prefer user B (to be recommended) having a decent \textit{occupation}; whereas user P who has a \textit{children} may dislike the \textit{drinking} or \textit{smoking} user Q. Thus, \textit{occupation} is a preferred attribute of user B to user A, and \textit{drinking} or \textit{smoking} is a \textit{repulsive attribute} of user Q to user P. Moreover, the \textit{salary} - \textit{occupation} forms a preference interaction between a pair of users, while \textit{children} - \textit{drinking} and \textit{children} - \textit{smoking} form the repulsiveness interaction. Obviously, different users may have different sets of preferred or repulsive attributes. 
Hereinafter, we call these attributes the key attributes to avoid ambiguity. 

To discover the key attributes, a simple solution is to enumerate all the attribute combinations, then measure the contribution of each combination to the reciprocal recommendation, and finally select the best set of attributes. Obviously, this solution is infeasible due to the exponential number of attribute combinations. 
Motivated by the aforementioned issues, in this paper we propose a reinforced random convolutional network (RRCN) approach, which can well capture the key attributes for reciprocal recommendation.
Particularly, we first develop an embedding component to capture the preferred and repulsive attributes from users' historical behaviors. Then, we build a feature embedding tensor between users' attributes and their preferred and repulsive attributes. Afterwards, we design a novel random CNN component, which performs a convolution operation on the feature tensor to capture the feature interactions. 
Different from conventional CNNs that can only convolute adjacent 
features, our proposed random CNN can randomly select features 
to convolute. 
We believe that by doing so, the convoluted features could well preserve feature interactions of key attributes. 
To further enhance the attributes selection process, we propose a reinforcement learning based strategy, which can select a set of salient attributes. Then for each user pair, we match both users' key attributes with the other users' attributes, based on which we make the reciprocal recommendation. 

In summary, our principle contributions are as follows:
\begin{itemize}
\item We propose a novel RRCN approach for reciprocal recommendation. To the best of our knowledge, this is the first attempt to perform reciprocal recommendation using the concept of key attributes and their interactions.
\item We propose a novel random CNN convolution operation method which could convolute non-adjacent features that are randomly chosen from the embedding feature tensor. Furthermore, a reinforcement learning based strategy is proposed to enhancing the attribute selection process by 
selecting salient attributes to form the candidate set of key attributes.
\item We evaluate RRCN on two real-world online dating datasets. The experimental results demonstrate that the proposed RRCN outperforms the state-of-the-art approaches in terms of several evaluation criteria.
\end{itemize}

%% file: preliminaries.tex
\section{Preliminaries}
\label{sec:problem}

As aforementioned, we model the 
reciprocal recommendation data as an attributed bipartite network 
$G$=$(U$=($M$, $F$), $E$, $A)$, where $U$ denotes the set of all the users including a subset $M$ of male users and a subset $F$ of female users, $E$ is the set of edges between female users and male users, and $A\in \mathbb{R}^{|U|\times L}$ is a set of attributes where $L$ is the number of attributes. Each user $u\in U$ is associated with an attribute vector $u_a\in \mathbb{R}^L \subset A$. For each directed edge $e$=$(m$, $f)\in E$, it means that a male user $m$ sent a message to a female user $f$. Note that if both edges ($m$, $f$) and ($f$, $m$) exist, then there is a reciprocal link between $f$ and $m$, denoted by $m\rightleftharpoons f$.

Meanwhile, for each male user $m$, we denote the set of female users by $P(m)$ that he has sent messages to, who are called preferred users of $m$. The set of female users who sent messages to $m$ but $m$ did not reply to them, called repulsive users of $m$, is denoted by $N(m)$.
Similarly, we use $P(f)$ and $N(f)$ to denote the sets of preferred and repulsive users of a female user $f$, respectively. 



\textbf{Problem definition.} Given a male user $m$ and a female user $f$ in the attributed bipartite network $G$, the reciprocal recommendation task is to develop a model, written as
\begin{equation}
\phi (G, m, f, \Theta),
\end{equation}
to accurately predict whether $m\rightleftharpoons f$ exists or not, where $\Theta$ represents the parameter setting of the model $\phi$.

Note that the output of $\phi$ falls in $[0,1]$ and a threshold is then used to determine whether a user should be recommended to another user or not.

%% file: approach.tex
\begin{figure*}[htbp]
    \centering
    \setlength{\belowcaptionskip}{5pt}
    \includegraphics[width=1\textwidth,height=2.8in]{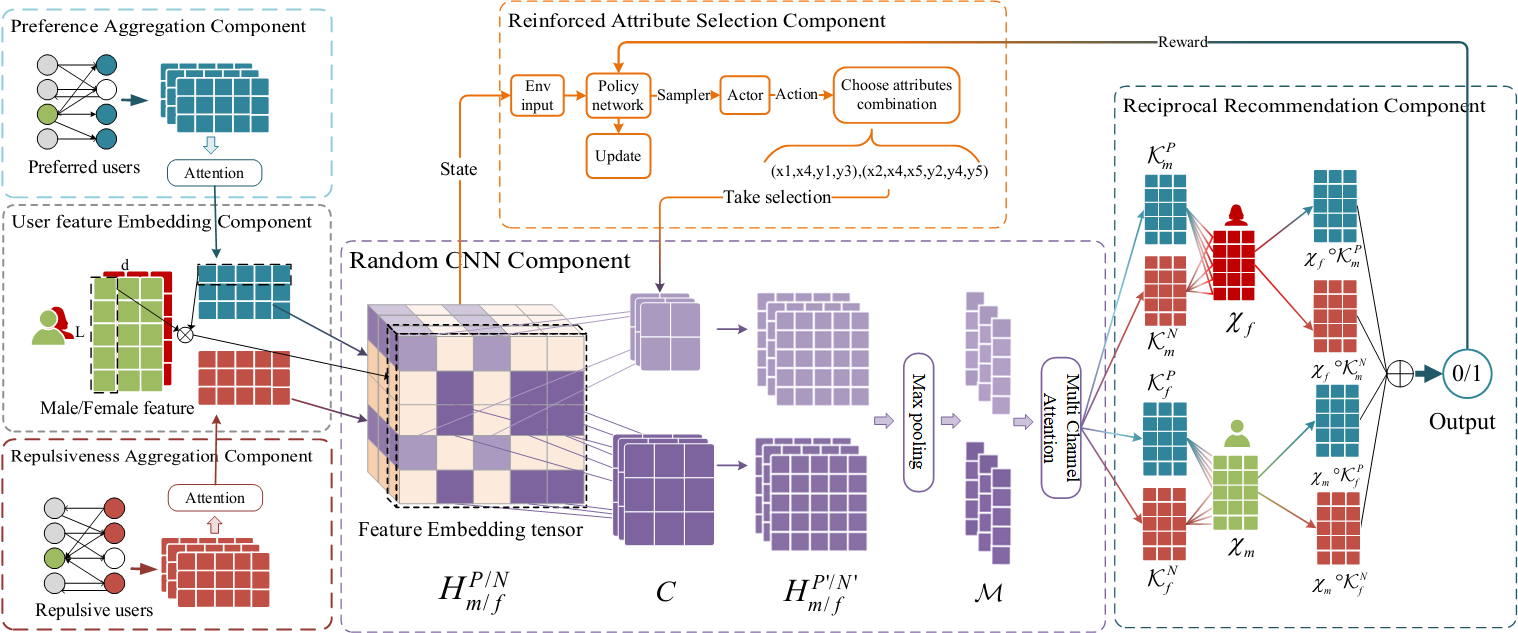}
    \caption{The overall framework of the proposed RRCN.}\label{fig:model}
    \vspace{-0.8cm}
\end{figure*}
\section{The Proposed RRCN}\label{sec:approach}

The framework of the proposed RRCN approach is depicted in Figure \ref{fig:model}, and it consists of four components: (1) user feature embedding component, (2) random CNN component, (3) reinforced attribute selection component and (4) reciprocal recommendation component. We detail each component in the following subsections. 
\vspace{-0.2cm}
\subsection{User feature embedding component}
This component is to embed users' attributes into a feature space. The working process is illustrated as follows. For a given male user $m$, we respectively extract his preferred user set $P(m)$ and repulsive user set $N(m)$ as highlighted in blue and red rectangles in Figure \ref{fig:model}.  The attributes of each user in $P(m)$ and $N(m)$ are embedded into a feature matrix denoted as $\chi \in \mathbb{R} ^{L\times d}$. Then, a soft-attention mechanism \cite{bahdanau2014neural} is employed to differentiate the importance of users in $P(m)$ and $N(m)$. The weight $\alpha_{i}$ of $i$-th user $u_i$ 
is calculated as
\begin{eqnarray} 
\setlength\abovedisplayskip{1pt}
\setlength\belowdisplayskip{1pt}
    \mu_{i} &=& \bm{W}_2^T \sigma_1(\bm{W}_1^T \chi'_{i} +b_1) + b_2 \\
    \alpha_{i} &=& \frac{e^{\mu_{i}}}{\sum_{u_j \in P(m),N(m)} e^{\mu_{j}}},
\end{eqnarray}

\noindent where $\sigma_1$ is $\tanh$ function, $\chi'_{i} \in \mathbb{R}^{Ld \times 1}$ is an one-dimension feature vector of user $u_i \in P(m)$ or $N(m)$ by a flattening operation, $\bm{W}_1^T \in \mathbb{R}^{Ld\times l_1}$ and $\bm{W}_2^T \in \mathbb{R}^{l_1\times 1}$ are neural network parameters. Then, the weighted feature representation $X_{P(m)}$ (of preferred users) and $X_{N(m)}$ (of repulsive users) 
is now calculated as
\begin{small}
\begin{equation}
    \setlength\abovedisplayskip{1pt}
    \setlength\belowdisplayskip{1pt}
\begin{split}
    X_{P(m)} &= \sum_{u_{i} \in P(m)} \alpha_{i} \chi_{i}\\
    X_{N(m)} &= \sum_{u_{i} \in N(m)} \alpha_{i} \chi_{i}
\end{split}
\end{equation}
\end{small}

Similar to xDeepFM, we respectively perform outer product operations between feature $\chi_m$ (of given user $m$) and $X_{P(m)}$ and $X_{N(m)}$,
along each embedding dimension. The output of this operation is a tensor denoted as $\bm{H}^P_{m},\bm{H}^N_{m} \in R^{L \times L \times d}$, 
written as
\begin{equation}
    \resizebox{0.8\columnwidth}{!}{$
    \begin{aligned}
    \bm{H}^{P}_{m} &= [{\chi_{m}}_1 \otimes {X_{P(m)}}_1,\cdots, {\chi_{m}}_d \otimes {X_{P(m)}}_d] \\ 
    \bm{H}^{N}_{m} &= [{\chi_{m}}_1 \otimes {X_{N(m)}}_1,\cdots, {\chi_{m}}_d \otimes {X_{N(m)}}_d]
\end{aligned}
$}
\end{equation}
Note that we have feature embedding tensor $\bm{H}^{P}_{m}, \bm{H}^{N}_{m}$ for a male user $m$ and $\bm{H}^{P}_{f}, \bm{H}^{N}_{f}$ for a female user $f$  by taking the same process as above. For simplicity reason, we denote these tensors using $\bm{H}^{P/N}_{m/f}$. 
This feature embedding tensor $\bm{H}^{P/N}_{m/f}$ is then fed into the next random CNN component. 
\vspace{-0.2cm}

\subsection{Random CNN Component}\label{sec:random}

\begin{figure}[htbp]
    \vspace{-0.2cm}
    \centering
    \includegraphics[width=1\columnwidth,height=3.3in]{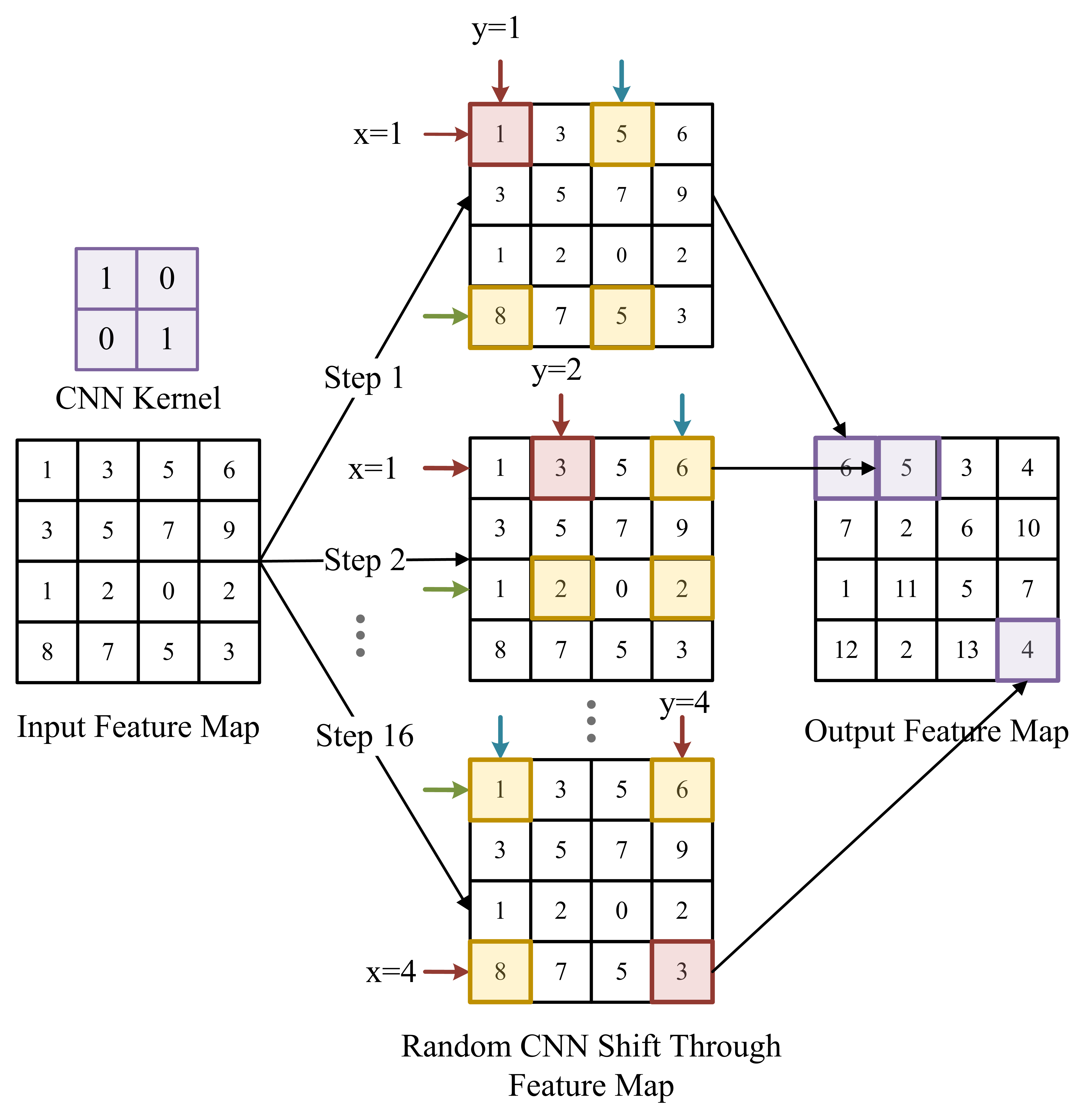}
    \caption{Illustration of attribute selection process of random CNN with kernel size as $2\times 2$. It traversely fixes each entry pointed by red arrows in the feature map, then randomly chooses the rest entries in a row-wise and column-wise manner to convolute. For instance, it fixes entry 1 at step 1, and then randomly chooses three entries to form a quad tuple (1,5,8,5) to convolute and generates entry 6 in the output feature map. It stops convoluting at step 16 as it has already traversed all entries contained in this feature map.} 
    \label{fig:randomcnn}
    \vspace{-0.3cm}
\end{figure}

In order to capture the key attributes and their interactions, a novel random convolutional operation is proposed to randomly convolute non-adjacent features. To convolute on a tensor $\bm{H}^{P/N}_{m/f}$, we define several kernels of different size to 
generate different attribute combinations. 
Then, the importance of these attribute combinations are learnt according to their contribution to reciprocal recommendation. The most important attributes are empirically considered as key attributes by this paper. An illustrating example of this random CNN is given in Figure \ref{fig:randomcnn}, and technical details of this component are illustrated as follows.


Let $k$ and $L$ respectively denote the number of key attributes and all attributes. Generally, we can enumerate all the attributes to build the candidate set of attribute combinations. However, the conventional CNN cannot convolute non-adjacent attributes, and thus cannot complete the enumeration process. To address this issue, we propose this random CNN component by revising the convolution operation to approximate the enumeration process. The size of convolutional kernel represents how many attributes should be convoluted. Given a $k\times k \times d$ kernel, the first row and column of this kernel is traversally fixed to an entry of $\bm{H}^{P/N}_{m/f}$. Then, we randomly select the rest $k-1$ rows and $k-1$ columns in $\bm{H}^{P/N}_{m/f}$, and the intersected matrix entries (of all $k$ rows and $k$ columns) form a k-sized feature tensor $\bm{H}_{xy} = [h_{ijl}]_{k\times k \times d}$ to convolute. By doing so, the complexity of random CNN operation is only $O(L^2)$ whereas the original complexity of enumeration is $O(C_L^k\times C_L^k)$, and thus we greatly reduce the computational cost.
The convolution operation over these selected attributes is calculated as,
\begin{equation}
    \setlength\abovedisplayskip{1pt}
    \setlength\belowdisplayskip{1pt}
    h^{'}_{xy} = \sum_{i=1}^{k} \sum_{j=1}^{k}\sum_{l=1}^{d}w_{ijl}h_{ijl}
\end{equation}
where $w_{ijl}$ is the weight of $h_{ijl}$. In the proposed random CNN component, we employ $K$ kernels of different size, i.e., $2 \times 2 \times d$, $3 \times 3 \times d$ and $4 \times 4 \times d$ where $d$ is the number of filters. Accordingly, a tensor $\bm{H}^{P'/N'}_{m/f}=[h'_{ijl}]_{L\times L \times d}$ is generated for $k$-sized kernels after the convolution operation.
Then, a max pooling layer \cite{graham2014fractional,tolias2015particular,nagi2011max} is applied on $\bm{H}^{P'/N'}_{m/f}$ in a row-wise manner, and it outputs a tensor $M^{P/N}_{m/f} \in \mathbb{R}^{L \times d}=[\rho_1,\cdots,\rho_L]$. This output of max pooling operation is also a feature vector representing interactive relationship among a set of $k$ key attributes. 

To recall that we have employed $K$ different kernels, and thus we have $K$ such feature vectors, denoted as $\mathcal{M}^{P/N}_{m/f} \in \mathbb{R}^{K \times L \times d}=[M_1, \cdots, M_K]$. To further differentiate the importance of each feature vector, a multi-dimension attention mechanism 
is proposed and calculated as
\begin{eqnarray}
\setlength\abovedisplayskip{1pt}
\setlength\belowdisplayskip{1pt}
    \mu_i &=& \sigma_1(M_i\bm{W}_d)\\
    \alpha_{i} &=& \frac{e^{\mu_{i}}}{\sum_{M_j \in \mathcal{M}} e^{\mu_{j}}}\\
    \mathcal{K}^{P/N}_{m/f} &=& \sum\limits_{M_i\in \mathcal{M}} \alpha_i M_i,
\end{eqnarray}
where 
$\bm{W}_d \in \mathbb{R}^{d\times 1}$ is the weight matrix of dimensions, $\alpha_i \in \mathbb{R}^{L\times 1}$ is the attention score of $M_i$, and $\mathcal{K}^{P/N}_{m/f}\in \mathbb{R}^{L\times d}$ is the aggregated feature embeddings of key attributes.
\begin{algorithm}[htbp]
    \caption{The Reinforced Random CNN Algorithm}\label{alg:rcnn}
    \renewcommand{\algorithmicrequire}{ \textbf{Input:}} 
    \renewcommand{\algorithmicensure}{ \textbf{Output:}} 
    \begin{algorithmic}[1]
        \Require { Feature embedding tensor $\bm{H}^{P/N}_{m/f}$, Kernel $C$}
        \Ensure {Convoluted feature embedding tensor $\bm{H}^{P'/N'}_{m/f}$}
        \Function{Convolution Oerpation}{$\bm{H}^{P/N}_{m/f}$,$k$}
            \For {$x$ $\leftarrow$ 0 to L}
                \For {$y$ $\leftarrow$ 0 to L}
                    \State RowIndex=$\{row_x\}\oplus SelectPolicy(x,k)$ 
                    \State ColumnIndex=$\{col_y\}\oplus SelectPolicy(y,k)$ 
                    \State 
                    \resizebox{0.74\columnwidth}{!}{
                    $h'_{xy}=\sum\limits_{i\in RowIndex}\sum\limits_{j\in ColumnIndex}\sum\limits_{l=1}^dw_{ijl}h_{ijl}$ }
                \EndFor
            \EndFor
        \State \Return {$\bm{H}^{P'/N'}_{m/f}$}
        \EndFunction
        \Function {SelectPolicy}{currentPosition,k}
            \State s=currentPosition 
            \State ActionsProbality=$P_{\theta}$($a_{xy} \mid s_{xy}, \bm{H}^{P/N}_{m/f}$)
            \State Actions=Sample(L,k-1,ActionsProbality)
            \State \Return{$Actions$}
        \EndFunction
    \end{algorithmic}
\end{algorithm}
\subsection{Reinforced attribute selection component}\label{sec:rl}

\begin{figure}[htbp]
    \centering
    \setlength{\belowcaptionskip}{1pt}
    \includegraphics[width=1\columnwidth,height=1.8in]{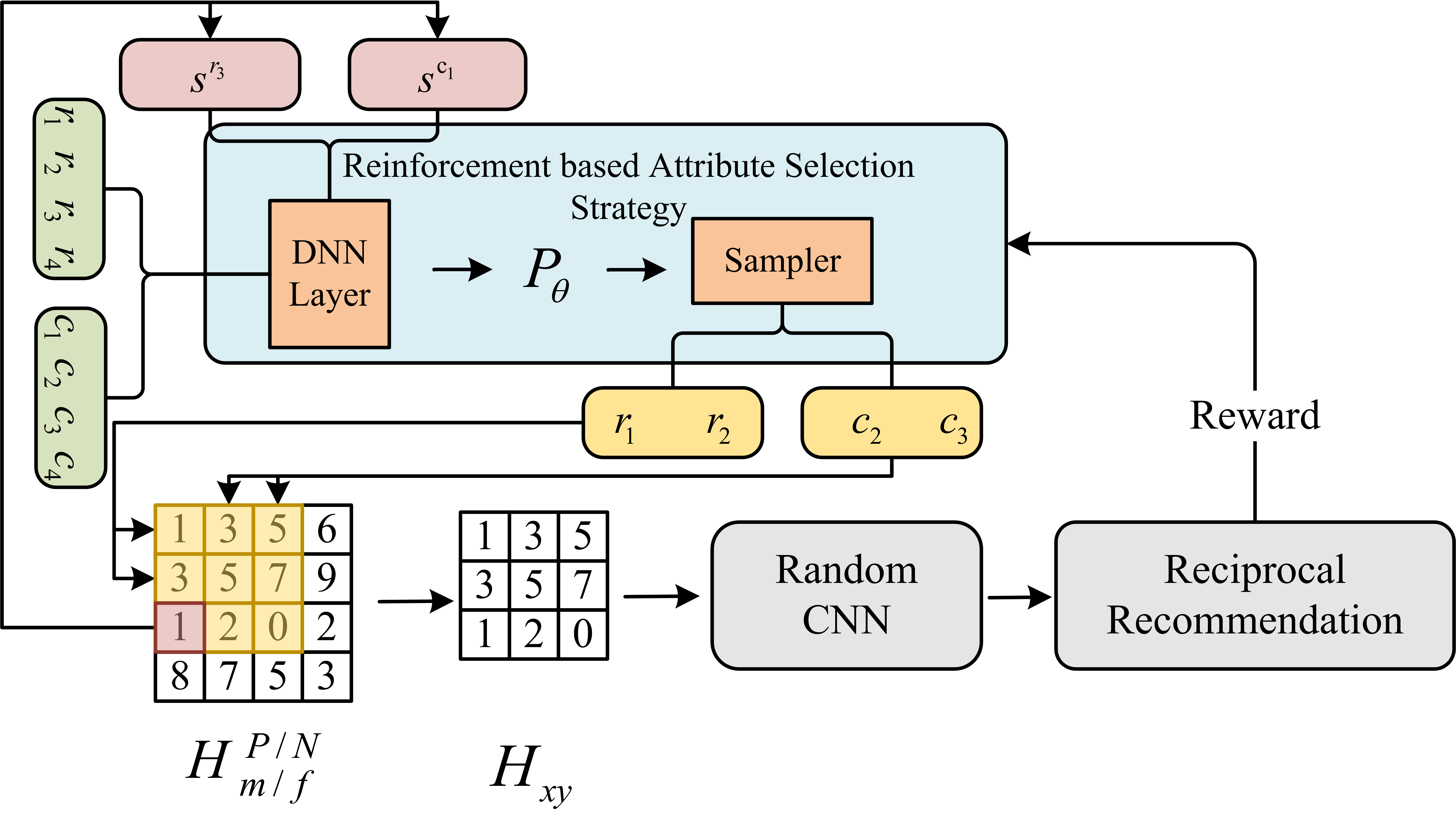}
    \caption{Illustration of the proposed reinforcement-enabled attribute selection strategy. In this figure, the red blocks are the initial states and the red cell of ``1'' is the fixed element to be interacted, the green blocks are candidate rows and columns, and the yellow blocks are indexes of selected rows and columns by taking an action.}
    \label{fig:reinforcelearning}
    \vspace{-0.6cm}
\end{figure}

To further enhance the feature selection process, a reinforcement learning \cite{kaelbling1996reinforcement,sutton1998introduction} based strategy is proposed to first select salient attributes as plotted in Figure \ref{fig:reinforcelearning}, and then apply the random CNN component to convolute these salient features.  

The proposed reinforced attribute selection component firstly fixes a cell $(x,y)$ as its initial state and takes action to choose the next $k^2-1$ entries to convolute, given a $k \times k$ kernel. Suppose the initial state $s_{xy}$ is set to the $x$-th row and $y$-th column, action $a_{xy} \in \mathcal{A}_k$ is to select next $k-1$ rows, i.e., $X=\{x_1,\cdots,x_{k-1}\}$, and next $k-1$ columns, i.e., $Y=\{y_1,\cdots,y_{k-1}\}$ from $\bm{H}^{P/N}_{m/f}$ to generate a submatrix $\bm{H}_{xy}=[h_{ijl}]_{k\times k\times d}$ for convolution. Its output is denoted as $s'_{xy}$. 
The probability of taking an action is approximated by a policy network $P_\theta(a_{xy}|s_{xy},\bm{H}^{P/N}_{m/f})$ consisting of two FC layers and a softmax layer written as,
\begin{eqnarray}
    \mu_{xy} &=& \bm{W}_2^T \sigma_1(\bm{W}_1^T \bm{H}^{P/N}_{m/f} +b_1) + s_{xy} \\
    P_{xy} &=& softmax(\mu_{xy}).
\end{eqnarray}
where $P_x=\{P(x_1),\cdots,P(x_L)\}$ and $P_y=\{P(y_1),\cdots,P(y_L)\}$ are the probability distributions of all the rows and columns.
Then, we sample $k-1$ rows and columns simultaneously according to their probability written as, 

\begin{equation}
    \begin{split}
    X &= Sample([x_1,\cdots,x_L],k-1,P_x)\\
    Y &= Sample([y_1,\cdots,y_L],k-1,P_y)
    \end{split}
\end{equation}

The reward of selecting attributes is estimated by their contributions to the model prediction accuracy, i.e., to minimize model loss, and thus the reward is calculated as 
\begin{equation}
    \resizebox{0.7\columnwidth}{!}{$
\begin{aligned}
            R(a_{xy},s_{xy},\bm{H}^{P/N}_{m/f}) = \sum_{x_i \in X} R(x_i, s_{xy}, \bm{H}^{P/N}_{m/f}) +\\ \sum_{y_j \in Y} R(y_j, s_{xy}, \bm{H}^{P/N}_{m/f})
\end{aligned} 
    $}
\end{equation} 
where
$R(x_i, s_{xy}, \bm{H}^{P/N}_{m/f})$ and
$R(y_j, s_{xy}, \bm{H}^{P/N}_{m/f})$ respectively denote the reward of choosing row $x_i$ and column $y_j$, 
written as 
\begin{equation}
    \resizebox{0.7\columnwidth}{!}{$
\begin{aligned}
    \setlength\abovedisplayskip{3pt}
    \setlength\belowdisplayskip{3pt}
    R(x_i,s_{xy},\bm{H}^{P/N}_{m/f}) &=& \sum\limits_{c=0}^{k-1}\sum\limits_{l=0}^{d} \frac{\partial{\mathcal{L}}}{\partial{s'_{xy}}} \frac{\partial{s'_{xy}}}{\partial{h_{icl}}} \\
    R(y_j,s_{xy},\bm{H}^{P/N}_{m/f}) &=& \sum\limits_{c=0}^{k-1}\sum\limits_{l=0}^{d} \frac{\partial{\mathcal{L}}}{\partial{s'_{xy}}} \frac{\partial{s'_{xy}}}{\partial{h_{cjl}}},
\end{aligned} 
    $}
\end{equation} 
where $\mathcal{L}$ is the model loss. The policy network is optimized by below objective function, 
given as 
\begin{equation}
    \resizebox{0.8\columnwidth}{!}{
    $\mathcal{J} =  \min \sum\limits_{(x,y)\in \bm{H}^{P/N}_{m/f}}\mathbb{E}_{a_{xy}} [ R(a_{xy},s_{xy},\bm{H}^{P/N}_{m/f})]$.
    }
\end{equation}
A policy gradient is calculated w.r.t. parameter $\theta$ using a widely adopted algorithm  \cite{williams1992simple,yu2017seqgan,wang2018neural}, and the corresponding gradient is directly given as
\begin{equation}
\resizebox{1\columnwidth}{!}{$
\begin{aligned}
    \mathbf{\bigtriangledown}_{\theta} \mathcal{J} &= \bigtriangledown_{\theta}
    \sum\limits_{(x,y)\in \bm{H}^{P|N}_{m/f}}\mathbb{E}_{a_{xy}} [ R(a_{xy},s_{xy},\bm{H}^{P/N}_{m/f})]\\
    &= \sum\limits_{(x,y)\in \bm{H}^{P/N}_{m/f}}\mathbb{E}_{a_{xy}} [ R(a_{xy},s_{xy},\bm{H}^{P/N}_{m/f})\bigtriangledown_{\theta} \log P_{\theta}(a_{xy}\mid s_{xy},\bm{H}^{P/N}_{m/f})].
\end{aligned} 
    $}
\end{equation} 
Then, the policy network is updated as $\theta\leftarrow\theta- \gamma\bigtriangledown_{\theta}\mathcal J(\theta)$.

\subsection{Reciprocal recommendation component}

This component is to predict whether a reciprocal link exits or not between any two users. Particularly, given a pair of users $(m, f)$, the feature embeddings of their key attributes could be calculated through previous components and are given as 
$\mathcal{K}_m^{P}$, $\mathcal{K}_m^{N}$, $\mathcal{K}_f^{P}$ and $\mathcal{K}_f^{N}$. 
Then, these features are concatenated as
\begin{equation}
        \nonumber \bm{V} = (\chi_m \circ \mathcal{K}_f^{P}) \oplus (\chi_m \circ \mathcal{K}_f^{N})\oplus (\chi_f \circ \mathcal{K}_m^{P}) \oplus (\chi_f\circ \mathcal{K}_m^{N}),
\end{equation}
where $\circ$ is vector dot product, and $\oplus$ denotes concatenation operation. This concatenated feature vector $\bm{V}$ is fed into two FC layers to make the reciprocal recommendation, and its model loss is designed as  

\begin{equation}
    \setlength\abovedisplayskip{1pt}
    \setlength\belowdisplayskip{1pt}
\resizebox{0.85\columnwidth}{!}{$
    \mathcal{L} = \min \sum\limits_{(m, f)\in U}-(y_{mf} \log(y'_{mf}) + (1-y_{mf})\log (1-y'_{mf}))
    $}
\end{equation}
\noindent where $y_{mf}$ is the true label whether the reciprocal link exists or not between $m$ and $f$, and we optimize the model 
using the Adam algorithm \cite{kingma2014adam}.

%% file: experiment.tex
\vspace{-0.3cm}
\section{Experiments}\label{sec:exp}

We perform extensive experiments on two real-world online dating datasets to answer the following research questions:
\begin{itemize}
    \item \textbf{Q1:} Does the proposed RRCN outperforms the state-of-the-art approaches for reciprocal recommendation task?
    \item \textbf{Q2:} How does CNN component and the reinforced learning based strategy affect the model performance?
    \item \textbf{Q3:} How does the reinforced random CNN capture the key attributes and their interactions?
\end{itemize}

\subsection{Datasets and evaluation criteria}
We consider two real-world online dating datasets  ``D1'' and ``D2''.
``D1'' is a public dataset provided by a national data mining competition \footnote{https://cosx.org/2011/03/1st-data-mining-competetion-for-college-students/}, which was originally collected from an online dating Website, and contains 34 user attributes and their message histories.
We use ``message'' and ``click'' actions between users to generate directed links between users.
``D2'' was collected by ourselves from one of the most popular online dating Websites \footnote{We anonymize the Website name for anonymous submission.}, which has over 100 millions of registered users, and each user has 28 attributes like age, marital status, constellation, education, occupation and salary.
We extract users who have sent or received more than 40 messages to build an attributed bipartite network, which consists of 228,470 users and 25,168,824 edges (each message corresponds to a directed edge).
The statistics of these two datasets are reported in Table \ref{tab:dataset}.

To evaluate the models, we adopt five popular evaluation metrics, i.e., Precision, Recall, F1, ROC, and AUC and the threshold is set to 0.5 for precision, Recall and F1. 

\begin{table}[htbp]
    \setlength{\abovecaptionskip}{10pt}
    \setlength{\belowcaptionskip}{5pt}
    \resizebox{\columnwidth}{!}{
    \begin{tabular}{cccp{2.5cm}<{\centering}p{1.2cm}<{\centering}p{1cm}<{\centering}}
        \toprule[2pt]
        Dataset        & \# Nodes    & \# Edges  & \# Reciprocal links & Max degree & Avg degree \\
        \hline
        D1 & 59,921  & 232,954     & 9,375 & 201,648 & 287          
        \\ \hline
        D2 & 228,470 & 25,168,824 & 1,592,945 & 4,231 & 110      \\ 
        \bottomrule[2pt]
    \end{tabular}
    }
    \caption{Statistics of experimental datasets.}
    \label{tab:dataset}
    \vspace{-0.5cm}
\end{table}

\vspace{-0.2cm}
\subsection{Baseline methods}
\label{sec:baseline}
As our task is a link prediction problem, and thus these top-K oriented reciprocal methods are not chosen for the performance comparison. In the experiments, we 
evaluate the proposed RRCN against the following feature embedding based approaches and link prediction approaches. 

\begin{itemize}
    \item \textbf{DeepWalk}~\cite{perozzi2014deepwalk} adopts the random walk to sample neighboring nodes, based on which nodes' representations are learned.
    
    \item \textbf{Node2vec}~\cite{grover2016node2vec} optimizes DeepWalk by designing novel strategies to sample neighbors.
    
    \item \textbf{DeepFM}~\cite{guo2017deepfm}  originally proposed for CTR prediction, is a factorization machine (FM) based neural network to learn feature interactions between user and item. 

    \item \textbf{xDeepFM} \cite{lian2018xdeepfm} uses multiple CIN components to learn high-order feature interactions among attributes. 
    \item \textbf{NFM} \cite{he2017neural} replaces the FM layer by a Bi-interaction pooling layer to learn the second order feature embedding.
    
    \item \textbf{AFM} \cite{xiao2017attentional} integrates the  FM layer with an attention mechanism to differentiate the importance of feature interactions.
    
    \item \textbf{DCN} \cite{wang2017deep} propose the deep cross network to capture the higher order feature interactions.
    
    \item \textbf{GraphSage} \cite{hamilton2017inductive} is an inductive graph neural network model, which generates the embedding of each node by randomly sampling and aggregating its neighbors' features.
          
    \item \textbf{PinSage} \cite{ying2018graph}, similar to the GraphSage, adopts the random walk to sample the neighbors of each node and aggregate them to represent the nodes feature.

    \item \textbf{Social GCN} \cite{wu2018socialgcn} is proposed to investigate how users' preferences are affected by their social ties which is then adopted for user-item recommendation task.  
\end{itemize}

\begin{table}[htbp]
    \vspace{-0.2cm}
    \setlength{\abovecaptionskip}{10pt}
    \setlength{\belowcaptionskip}{5pt}
    \centering
    \Huge
    \resizebox{\columnwidth}{!}{
    \begin{tabular}{ccccccc}
        \toprule[2pt]
        \multirow{2}*{\textbf{Methods}}                         & \multicolumn{3}{c}{\textbf{D1}}
                                                                & \multicolumn{3}{c}{\textbf{D2}} \\
        \cmidrule(lr){2-4}
        \cmidrule(lr){5-7}
                                                                & Precision       & Recall          & F1         & Precision       & Recall          & F1                                \\
        \midrule
        DeepWalk                                                & .5177          & .3544          & .4208          & .8801          & .7579          & .8144          \\
        Node2vec                                                & .4865          & .4380          & .4610           & .8138          & .8558          & .8343          \\ \hline
        DeepFM                                                  & .7004          & .4852          & .5732          & .8533          & .7477          & .7970          \\ 
        xDeepFM                                                 & \underline{.7714}          & .5094          & .6136          & \underline{.9357}          & \underline{.8605}          & \underline{.9047}          \\ 
        NFM                                                     & .5685          & .5593          & .5639       &  .7856           & .8198           & .8024\\
        AFM                                                     & .4210          & .3733          & .3957        & .7997           & .8214           & .8104\\
        DCN                                                     & .5143          & .5566          & .5346        & .7896           & .8436          & .8157\\
        
        \hline
        GraphSage                                               & .7151          & .3383          & .4593          & .6829          & .6643          & .6735          \\
        PinSage                                                 & .6428          & \underline{.7493}          & \underline{.6920}         & .7220          & .7549          & .7991          \\
        SocialGCN                                               & .4667          & .4434          & .4547          & .8588          & .8314          & .7991          \\ \hline
        RRCN                                                    & \textbf{.7865} & \textbf{.7695} & \textbf{.7779} & \textbf{.9686} & \textbf{.8679} & \textbf{.9154} \\
        \bottomrule[2pt]
    \end{tabular}
    }
    \caption{Results of reciprocal recommendations on ``D1`` and ``D2''.}
    \label{tab:DatingResult}
\end{table}

\begin{figure}[htbp]
\centering
\vspace{-0.5cm}
    \begin{minipage}{.49\columnwidth}
        \includegraphics[width=1\columnwidth,height=1.5in]{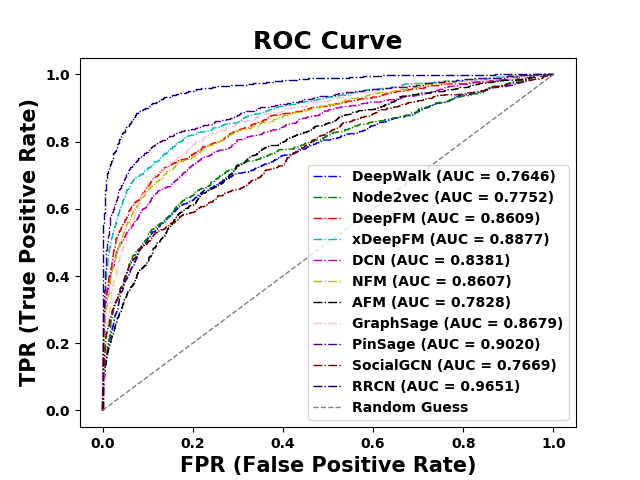}
    \caption{ROC results of all approaches on ``D1``.}
    \label{fig:roc_sjjy}
    \end{minipage}
    \begin{minipage}{.49\columnwidth}
        \includegraphics[width=1\columnwidth,height=1.5in]{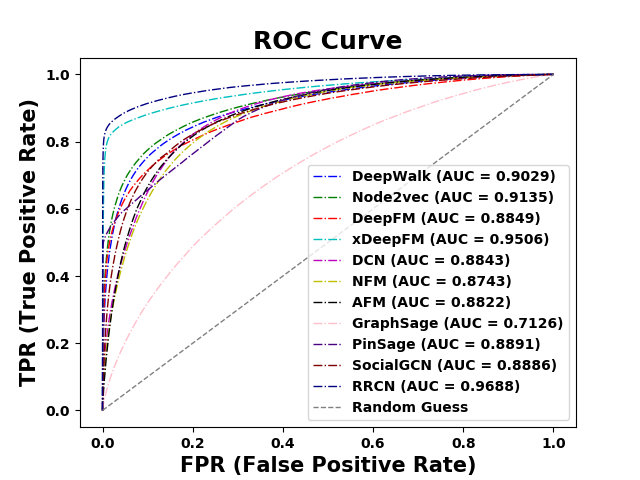}
    \caption{ROC results of all approaches in ``D2 ''.}
    \label{fig:roc}
    \end{minipage}
    \vspace{-0.3cm}
\end{figure}

\subsection{Results on reciprocal recommendation (Q1)}
\label{sec:Q1}

This experiment is to verify whether RRCN outperforms the state-of-the-art approaches for the reciprocal recommendation task.
Before the experiments, we first extract all the reciprocal links and negatively sample the same number of non-reciprocal links from the two datasets which are randomly partitioned into training data and testing data at the ratio of 80\% to 20\%, respectively.
Afterwards, we run all the comparison models on all the datasets and report the experimental results in Table \ref{tab:DatingResult}, Figure \ref{fig:roc_sjjy}, and \ref{fig:roc}, respectively. 

Table \ref{tab:DatingResult} shows the results on precision, recall, and F1-score. We can see that RRCN consistently outperforms other approaches.
We can also see that feature embedding based approaches, i.e., xDeepFM could achieve better performance than other baseline models. This is consistent with our common sense that users' attributes play a more important role in reciprocal recommendation. Nevertheless, these approaches convolute all attributes which in turn generates unnecessary information, and thus deteriorates the model performance.
Besides, graph representation learning based approaches, i.e., PinSage, GraphSage and SocialGCN, achieve better performance on ``D1'' which is a smaller dataset, but are the worst on a larger dataset. This implies that these approaches are good at capturing graph structural features but need to design a better manner to combine users' attributes and interactive behavior features.

Figures \ref{fig:roc_sjjy} and \ref{fig:roc} respectively plot the AUC results on both datasets, where the X-axis of ROC is FPR (false positive rate) indicating the rate that a classifier wrongly classifies false data, and Y-axis of ROC is TPR (true positive rate) indicating the rate that a classifier correctly labels true data. Obviously, it is desired to see a ROC curve having a higher TPR value and a lower FPR value at the same time, and such curve also has a larger AUC value. From the figures, we can see that RRCN achieves the highest AUC (0.9651 and 0.9688) respectively on ``D1'' and ``D2''.

In summary, we conclude that our proposed RRCN achieves the superior performance against a number of SOTA 
approaches in terms of five evaluation criteria.

\begin{table}[htbp]
    \vspace{-0.1cm}
    \setlength{\abovecaptionskip}{10pt}
    \setlength{\belowcaptionskip}{5pt}
    \centering
    \Huge
    \resizebox{1\columnwidth}{!}{

    \begin{tabular}{ccccccc}
        \toprule[2pt]
        \multirow{2}*{\textbf{Kernel size}} &
        \multicolumn{4}{c}{\textbf{Model performance}}                                                                                             \\
        \cmidrule(lr){2-6}
                                            & Precision          & Recall             & F1                 & ACC                & AUC                \\
        \midrule
        CCNN+K=2           & .7400             & .5066             & .3482             & .5067             & .8530             \\
        CCNN+K=3           & .7093             & .5746             & .4931             & .5747             & .8001             \\
        CCNN+K=4           & .7057             & .5270             & .3957             & .5270             & .7876             \\
        \hline
        DCNN+K=2+D=2             & .7437             & .5606             & .4590             & .5607             & .8600             \\
        DCNN+K=2+D=2             & .8161             & .8140             & .8150             & .8140             & .8688 \\
        DCNN+K=2+D=2             & .6319             & .6315             & .6312             & .6315             & .6834             \\
        \hline
        RCN+K=2          & .8180/.17        & .7751/.25        & .7422/.32        & .7751/.25        & .8759/.15       \\
        RCN+K=3           & .7497/.19        & .7159/.24        & .6889/.29        & .7159/.24        & .7967/.20        \\
        RCN+K=4          & .7655/.19        & .6985/.29        & .6455/.19        & .7654/.40        & .8281/.17        \\
        \hline
        RRCN+K=2                         & \underline{.9181}             & \underline{.9172}             & \underline{.9171}             & \underline{.9172}             & \textbf{.9702}    \\
        RRCN+K=3                         & \textbf{.9249}    & \textbf{.9190}    & \textbf{.9187}    & \textbf{.9190}    & \underline{.9689}             \\
        RRCN+K=4                         & .8720             & .8632             & .8624             & .8632             & .9634             \\
        \bottomrule[2pt]
    \end{tabular}
    }
    \caption{Experimental results of the proposed random CNN and the reinforcement learning enabled random CNN on ``D2''.}
    \label{tab:CNNCompare}
    \vspace{-0.7cm}
\end{table}
\begin{table*}
\centering
\resizebox{0.8\textwidth}{!}{
\begin{tabular}{m{1.2cm}<{\centering}m{1.2cm}<{\centering}p{6cm}<{\centering}p{6cm}<{\centering}} 
\toprule[2pt]
\textbf{Kernel Size} & \textbf{Method} & \textbf{Initial State}  & \textbf{Final State} \\
\hline
\multirow{2}*{K=2} & CNN & \{(Education,Occupation), (Salary,Smoking)\}   & \{(Education,Occupation), (Salary,moking)\} \\
\cline{2-4}
  & RRCN & \{(Education), (Salary)\}   & \{(Education,\textbf{Salary}), (Salary,\textbf{House})\} \\
\hline
\multirow{2}*{K=3} & CNN & \{(Education,Occupation,Salary), (Salary,Smoking,Drinking)\}   & \{(Education,Occupation,Salary), (Salary,Smoking,Drinking)\} \\
\cline{2-4}
  & RRCN & \{(Education), (Salary)\}   & \{(Education,\textbf{Occupation},\textbf{House}), (Salary,\textbf{Occupation},\textbf{Education})\} \\
\hline
\multirow{2}*{K=4} & CNN & \{(Education,Occupation,Salary,Smoking), (Salary,Smoking,Drinking,Children)\}   & \{(Education,Occupation,Salary,Smoking), (Salary,Smoking,Drinking,Children)\} \\
\cline{2-4}
  & RRCN & \{(Education), (Salary)\}  & \{(Education,\textbf{Salary},\textbf{House},\textbf{Children}), (Salary,\textbf{Height},\textbf{Occupation},\textbf{Hometown})\} \\
\bottomrule[2pt]
\end{tabular}
}
\caption{Results on preferred attributes selected by conventional CNN and RRCN. The bold features in final state indicate that they are selected by taking an action given initial state in RRCN.}\label{tab:featureRL}
\vspace{-0.5cm}
\end{table*}
\subsection{Effect of random CNN component and reinforced feature selection strategy (Q2)}
\label{sec:Q2}

In this experiment, we perform an ablation study to evaluate the effect of both random CNN operations (denoted as RCN) and reinforcement learning based strategy (denoted as RRCN). 
We also compare the model performance by replacing the random CNN with conventional CNN (\textbf{C}CNN) and dilated CNN (\textbf{D}CNN).
Note that for lack of space, we only show the results on the larger dataset ``D2''.

For all approaches above, the kernel size (\textbf{K}) is respectively set to 2, 3 and 4. The results are reported in Table \ref{tab:CNNCompare}. Clearly, the performance of conventional CNN with different kernel size is the worst, as shown in the first three rows. The dilated CNN could be considered as a special case of our approach. We set dilation rate (\textbf{D}) to 2 for all experiments.
The performance of dilated CNN is better than that of the conventional CNN, and this verifies our assumption that the convolutions of non-adjacent features could enhance model prediction ability.
On average, our proposed random CNN component is better than all compared methods. However, the performance of random CNN component is not stable, as shown by its mean value and standard variance value of 5 results.
Moreover, we can see that RRCN achieves the best performance on all the evaluation criteria. Particularly, the performance of ``RRCN+K=3'' achieves the best results, where ``K=3'' means that three key attributes should be convoluted.
From this result, we can infer that a combination of three attributes is able to capture salient preferred or repulsive attributes and their feature interactions.

\vspace{-0.2cm}
\subsection{A case study on how RRCN captures the key attributes and their interactions (Q3)}\label{sec:Q3}

To further show the effect of the reinforcement learning based strategy, we report intermediate results of preferred features selected by RRCN in Table \ref{tab:featureRL}.
Specifically, we first fix the initial cell in the feature matrix to \textit{(Education, Salary)} which indicates the attributes of male and his preferred users are Education and Salary, respectively.
Then, we report the initial state and final state for both conventional CNN and RRCN.
Note that conventional CNN simply slides adjacent $k\times k$ features in the feature map, and thus its initial and final states are determined by the sequence order of features in the feature matrix.
For RRCN, it takes an action through the designed RL strategy, and the selected features by an action are highlighted in bold as reported in the final state.
For $k$=3, the final state of CNN is \textit{\{(Education, Occupation, Salary), (Salary, Smoking, Drinking)\}} for a user and the preferred attribute interaction tensor to convolute. Clearly, the male user has some undesired attributes like Smoking and Drinking, and thus the output of the convolution may not contribute to the final recommendation.
For RRCN, the final state is \textit{\{(Education, Occupation, House),(Salary, Occupation, Education)\}}. Obviously, the RRCN can select more preferred attributes of the user based on the interactions between the preferred attributes and user's own attributes.
For $k>3$, it may not be able to find a more suitable attribute, as shown in final state, to join the combination, and thus the model performance will not further increase. This further verifies the merit of the proposed RRCN.

%% file: relatedwork.tex
\vspace{-0.2cm}
\section{Related Work}\label{sec:related}
The reciprocal recommendation has attracted much research attention \cite{brozovsky2007recommender,akehurst2011ccr,li2012meet,xia2015reciprocal,wobcke2015deployed,vitale2018online,xia2019we}.
In \cite{brozovsky2007recommender}, a collaborate filtering (CF) based approach is proposed to compute 
rating scores of reciprocal users. 
The proposed RECON \cite{pizzato2010recon} considers mutual interests between a pair of reciprocal users. 
Alternatively, \cite{xia2015reciprocal} calculates both the reciprocal interest and reciprocal attractiveness between users.
\cite{vitale2018online} designs a computationally efficient algorithm that can uncover mutual user preferences. 
\cite{kleinerman2018optimally} proposes a hybrid model which employs deep neural network to predict the probability that target user might be interested in a given service user.
However, these approaches mainly consider the preferred attributes, but overlook the repulsive attributes. Moreover, they treat all attributes equally, which ignores the fact that different attributes may have different impacts on the reciprocal recommendation, and this partially motivates our work. 

Essentially, our proposed approach is feature embedding based approach \cite{shan2016deep,zhang2016deep,qu2016product,cheng2016wide}. Among the feature embedding based approaches \cite{he2017neural,xiao2017attentional,zhou2018deep,zhou2019deep}, the SOTA DeepFM~\cite{guo2017deepfm} extracts both first and second order feature interactions for CTR problem, while xDeepFM \cite{lian2018xdeepfm} further employs multiple CINs to learn higher order feature representation. 
As aforementioned, 
this paper technically designs a random CNN component, by convoluting non-adjacent attributes, to approximate the enumeration of all attribute combinations to discover key attributes. 
Bearing similar name to ours, the random shifting CNN \cite{zhao2017random} designs a random convolutional operation by moving the kernel along any direction randomly chosen from a predefined direction set. 
However, this model still convolutes 
adjacent features. The dilated CNN \cite{yu2015multi} can convolute non-adjacent features but it only convolutes features spanning across a fixed interval which might miss some attribute combinations. However, our proposed approach randomly (or based on a reinforced strategy) chooses the intersections of rows and columns from the feature interaction matrix to convolute 
the non-adjacent features, which is our major technical contribution to the literature. 

%% file: conclusion.tex
\vspace{-0.2cm}
\section{Conclusion}
\label{sec:con}

In this paper, we propose a novel reinforced random convolutional network (RRCN) model for reciprocal recommendation task. First, we assume that a set of key attributes as well as their interactions are crucial to the reciprocal recommendation. To capture these key attributes, we technically propose a novel random CNN operation method which can randomly choose non-adjacent features to convolute. To further enhance this attribute selection process, a reinforcement learning based strategy is proposed. Extensive experiments are performed on two real-world datasets and the results demonstrate that RRCN achieves the state-of-the-art performance against a number of compared models. 
